\documentclass[sort&compress ,final] {aipproc}
\usepackage{sidecap,graphicx,boxedminipage,epsfig,wrapfig,floatflt,shadow}

\layoutstyle{8x11double}

\begin{document}

\title{
Increasing interest and awareness about teaching in science undergraduates
}

\classification{01.40Fk,01.40.gb,01.40G-,1.30.Rr}
\keywords      {physics education research}

\author{C. Singh, L. Moin, and C. Schunn}{
address={Department of Physics $\&$ Astronomy and Learning Research $\&$ Development Center, 
University of Pittsburgh, Pittsburgh, PA, 15260}
}

\begin{abstract}
We discuss the development, implementation, and assessment of a course for science undergraduates designed to
help them develop an awareness and a deeper appreciation of the intellectual demands of physics teaching.
The course focused on increasing student enthusiasm and confidence in teaching by providing well supported teaching opportunities
and exposure to physics education research. The course assessment methods include 1) pre/post-tests measures of
attitude and expectations about science teaching, 2) self and peer evaluation of student teaching, 3) content-based
pre/post-tests given to students who received instruction from the student teachers, and 4) audio-taped focus group
discussions in the absence of the instructor and TA to evaluate student perspective on different aspects of the
course and its impact.
\end{abstract}

\maketitle

\section{Background}

In the past four decades, the science and math K-12 teaching field has witnessed a marked 
decline in recruitment and the number of undergraduates pursuing math and science 
teaching careers has dropped from $22\%$ in 1966 to $5\%$ in 1988~\cite{green}.  
In the U.S., there is a serious shortage of qualified math and science teachers~\cite{villegas,forcier,gafney,shugart}.
Nationally, from 1993 to 1999, $39\%$ of school districts reported math and science teacher 
vacancies, and $19\%$ reported difficulty filling those vacancies~\cite{usdoe}.
According to a nationwide survey released by the National Science Teachers Association 
in 2000, this trend continues and $61\%$ of high schools and $48\%$ of middle schools experienced 
difficulty locating qualified science teachers to fill vacancies. The situation is expected to worsen given 
that over $30\%$ of current teachers with varying years experience report considering leaving the 
profession~\cite{nsta}.  Nearly every region of the United 
States reported considerable teacher shortages, with most prominent
need for mathematics teachers closely followed by those in physics, chemistry, earth/physical 
science, and biology~\cite{aaee}.  
The problem is so pronounced that since the early 1990s, in-need districts across the nation have imported 
thousands of international K-12 math and science teachers~\cite{brulliard}.

Numerous remedies have been attempted to alleviate this teaching shortage. Remedies 
range from national to local policies and programs and include such approaches as emergency 
certification and out-of-field assignment to fill vacancies; alternative certification programs to 
hasten licensing requirements and job placement; tapping nontraditional candidate pools such as 
paraprofessionals, retired military, or career changers; providing scholarships, signing bonuses, 
or student loan forgiveness; and establishing partnerships between school districts and teacher 
preparation institutions to meet staffing needs cooperatively~\cite{aaee,forcier,usdoea}.
Each remedy has certain costs and some degree of 
success. However, many remedies must resort to back peddling to meet content knowledge 
qualifications, calling back to the educational fold those who have already left, or investing in 
populations who fail to complete licensing requirement. 

\section{Introduction}

One of the most accessible potential recruits are science and engineering undergraduates who have not yet 
completed a degree. 
According to a longitudinal research study conducted by Seymour and Hewitt~\cite{seymore},  $20\%$ of SEM undergrads 
at one time consider careers in math or science teaching, although less than $8\%$ of them hold to the career 
interest. We developed, implemented and assessed a course called ``Introduction to physics teaching" for 
science and engineering undergraduates so that they would consider K-12 teaching as a potential career choice.
The course was designed to increase awareness and develop a deeper appreciation
about the intellectual demands of physics teaching. The course attempted to increase student enthusiasm and confidence
in teaching by giving them opportunity to design instructional modules in pairs and teach in authentic college
recitation classes two times during the semester. 
We provided significant scaffolding support and guidance during the development of the modules but gradually
decreased the guidance to ensure that students develop confidence and self-reliance.
The course strived to improve students' knowledge of 
pedagogical issues, familiarize them with cognitive research and its implication
for teaching physics, and included extensive discussions of physics education research including topics
related to active engagement, effective curricula, student difficulties in learning
different physics content, epistemological and affective issues. Special attention was paid to helping
students see the relevance of these discussions to actual classroom teaching.

\section{Course Details}

The course has been taught twice with a total of 12 students. A majority of students were science and
engineering undergraduates (sophomores-seniors) with two masters students from the school of education.
The cumulative grade point average for the students was between 2.5 to 3.5. At least a B grade average in 
introductory physics I and II was mandatory to enroll which was a requirement imposed by the department of
physics because each student pair was required to conduct two college recitations.

An initial survey in the first class period to the students enrolled in the class suggests that a majority 
of students had some kind of teaching experience. The most common teaching experience was tutoring in high
school. The survey responses suggest that students felt confident in teaching the subject matter they tutored
earlier. When asked to rank-order the main reasons for having taught in the past, the students cited ``curiosity"
followed by ``a sense of being good at it", followed by ``a desire to work with children", and
``giving back to the community".

The class met for 3 hours per week for a semester and students obtained 3 credits for it.
Students were assigned readings of one or two journal articles about teaching and learning each week. 
They submitted answers to the questions assigned about the readings and discussed
the articles in class each week.
We used a field-tested ``Cognitive Apprenticeship Model"~\cite{cog} of teaching and learning which has three major
components: modeling, coaching, and fading. Modeling in this context refers to the instructor demonstrating
and exemplifying the criteria of good performance. Coaching refers to giving students opportunity to
practice the desired skills while providing guidance and support and fading refers to weaning the support
gradually so that students develop self-reliance. In the modeling phase, students worked through and
discussed modules from an exemplary curriculum, Physics by Inquiry~\cite{inquiry}, in pairs. There was extensive
discussion of the aspects of the modules that make them effective and the goals, objectives, and performance
targets that must have been outlined before those modules were developed. In the coaching and fading phases,
the student pairs developed, implemented and assessed two introductory physics tutorials and related pre-/post-tests
with scaffolding support from the instructor, teaching assistant (TA) and peers. 

Students were allowed to choose their partner and they stayed with the same partner for both tutorial.
All student pairs designed the two tutorials on the same broad topics: DC circuits and electromagnetic
induction. Although all student pairs employed the tutorial approach to teaching, there was flexibility in how to design the
tutorial.
For example, one group successfully employed cartoons in their tutorials.
Also, students were free to choose the focus of their 25 minute long tutorial (10+15 minutes
were spent on the pre-test and post-test respectively). Each student group determined the goals and performance
targets for their tutorial which was then discussed during the class. This class discussion was very useful in
helping students realize that they needed to sharpen their focus for a 25 minute tutorial instead of covering every
concept in DC circuits or electromagnetic induction. A majority of the preliminary development of the tutorials and the 
accompanying pre-/post-tests took place outside of the class and students iterated versions of the tutorials
with the instructor and TA. Then, each pair tested their pre-/post-tests and tutorials on fellow classmates and
used the discussion and feedback to modify their tutorial. The peers were very conscientious about providing
comments on both the strengths and weaknesses of the tutorials.

\section{Course Evaluation}

The content-based pre-/post-tests accompanying the tutorials were given to the introductory physics students during the
recitation. The typical pre-/post-test scores were $40\%$ and $90\%$ respectively with a Hake normalized gain of 
$0.8$~\cite{hake}. We note that the pre-test refers to the test given after traditional classroom instruction but before the
tutorials.

We developed a teaching evaluation protocol which includes 15 questions on a Likert scale designed to evaluate different
aspects of teaching. The 15 questions in the protocol were further divided into two parts: the first 7 questions were
related to content/lesson plans/class design and the other 8 questions dealt with the class activities during instruction.
The following are some items:
\begin{itemize}
\item Class content was designed to elicit students' prior knowledge and preconceptions and build new concepts from there.
\item The lesson was designed to engage students as members of a learning community: engaged in talk that builds on each other's
ideas, that is based on evidence and responds to logical thinking.
\item Instructional strategy included useful representational tools (for example, symbols, charts, tables, and diagrams).
\item The activity actively engaged and motivated students rather than having them be passive receivers.
\end{itemize}
Each student was required to observe and critique the instruction of at least one other
pair in each of the two rounds in addition to evaluating their own performance. All of the classroom teaching by the
students were video-taped. After each round, we discussed the teaching evaluations of each group in class to stress
the aspects of teaching that were good and those in need of improvement. We found that the student evaluation of other
pairs were quite reliable and consistent with the instructor and TA evaluation. Students did a good job evaluating
the positive and negative aspects of other group's instruction. However, self-evaluations were not reliable and students
always rated themselves highly. Students were told that their grades will depend only on the evaluation
conducted by the instructor and the TA and not on the self and peer evaluations and that the self and peer evaluations
were to help them learn to critique various aspects of instruction. The fact that students rated themselves
higher than others may be because they were worried that the evaluation may factor into their course grade.

There was a clear difference between different student pairs in terms of how effectively they helped the introductory
physics students work on the tutorials in groups. There was a strong correlation between the extent to which group work
was motivated and emphasized at the beginning of the recitation and its benefits explained and whether introductory students
worked effectively in groups. There were discussions of these issues with the student pairs and each student pair
obtained a copy of all of their evaluations. They were asked to pay attention to the instructor/TA/peer critiques of their
performance. However, the second performance of each pair was not significantly different from the first. For example,
pairs good at employing group work effectively the first time did it well the second time and those who
had difficulty the first time had similar difficulties the second time. More detailed guidance is needed for 
improving students' classroom delivery methods.

We also conducted an anonymous survey in the absence of course instructor at the end of the course. One of the questions
on the survey asked students to rate how the course affected their interest in becoming a teacher. $56\%$ reported a significant
positive impact, $34\%$ a positive impact and $10\%$ no impact. Students noted that they learned about the intellectual rigor of
instructional design from moderate to great extent. On a scale of 1 to 5, students were asked to rate different elements that
contributed to learning. They provided the following responses:
\begin{itemize}
\item Preparing tutorials and presentations: 4.8/5
\item Instructor's feedback on these: 4.5/5
\item Class discussions: 4.3/5
\item Rehearsals for their presentation: 4/5
\item Instructor's presentations: 4/5
\item readings: 3.9/5
\end{itemize}

We also conducted an audio-taped focus group discussion to obtain useful feedback to evaluate and improve next offering of the
course. The focus group was conducted on the last day of class in the absence of the instructor and the TA. The facilitator
asked students pre-planned questions for one hour. The questions and some typical responses are presented below:

\underline{Question 1:} What is the take home message of this course?
\begin{itemize}
\item S1: Teaching is more than teacher's perception. How much of a two way relation is necessary to teach students.
\item S2: Helped me understand that teachers have to learn from students.
\item S3: Instruction is more students. There are methods available to make instruction more suited to students. There is
a mountain of cognitive research that is being developed as a resource for me as a future teacher...that was my biggest
fear when we started talking about bringing instruction to student's level.
\item S4: Increased enthusiasm. You have to take into account student's level.
\item S5: Increased appreciation of teaching. Opened my eyes to the difficulty and different techniques for teaching students 
with different prior knowledge.
\item S6: Figuring out different ways of making students active and structuring the lessons so that there is a lot of activity
by students to learn on a regular basis.
\end{itemize}
\underline{Question 2:} Do you take a different perspective during your own classes after you learned something about how to teach?
\begin{itemize}
\item S1: I think now that teachers who don't teach well could be trained but before the course I just took it for granted that
there are good and there are bad teachers and that's all.
\item S2: My college instructors ignore the work being done in how people learn.
\item S3: Slightly, because I know how difficult it is. I give more respect to good teachers.
\item S4: It gives you an idea about how a teacher cares about the students.
\end{itemize}
\underline{Question 3:} What did you learn from your K-12 teaching? How do you compare that to teaching at the college level?\\
A common response was that the students had not thought about teaching issues in high school or till they took this course.
\begin{itemize}
\item S1: When I was a student I just took teaching for granted and did what they told me to.
\item S2: I never thought about teaching when I was in high school.
\item S3: At school most were educators in college not.
\end{itemize}
\underline{Question 4:} How did this course affect your interest in teaching? Tell me about your plans for pursuing teaching.\\
All students except two said they will teach. A majority explicitly said they plan to teach in high school. 
\begin{itemize}
\item S1: Reinforces my interest. Made me realize that I don't want to teach college because of the structure of college-lots of
material, little support, under-appreciated...I want to have more time to engage students in the method learned in this course. 
\item S2: It helped me decide I want to go on to teaching right after college.
\item S3: I want to be a teacher. This course affected me positively.
\item S4: K-12. Good physics teacher in high school to give good base at young age...early
\end{itemize}
\underline{Question 5:} How could this course be improved to enthuse more people to teaching?\\
One common discouraging response was that students felt they did not really get an opportunity
to teach where the word teaching referred to frontal teaching. Despite the fact that the course
attempted to bridge the gap between teaching and learning, students felt that moving around the
classroom helping students while they worked on the tutorials was not teaching.
Common suggestions included a follow-up class with the following features:
\begin{itemize}
\item Observing, critiquing $\&$ delivering frontal teaching
\item Observing and critiquing K-12 teaching
\item Amount of reading per week can be reduced although students appreciated the readings
\end{itemize}

\section{Summary}

We developed, implemented and assessed a course for science undergraduates to increase their interest and awareness
about teaching issues. In addition to extensive discussions about issues related to teaching and learning, student
pairs designed and implemented two tutorials in college recitation classes. Assessment methods include pre-/post-tests
of expectation and attitude about teaching, content-based pre-/post-tests before and after tutorials designed by students,
critiquing peer and self-evaluation of teaching and focus group discussions.



\bibliographystyle{aipproc}  
{}
\end{document}